\documentclass[twocolumn, prl,
 amsmath,amssymb,
 superscriptaddress,aps,
]{revtex4-2}
\usepackage[utf8]{inputenc}
\usepackage[english]{babel}
\usepackage{amsmath}
\usepackage{hyperref}
\usepackage{amsfonts}
\usepackage{amssymb}
\usepackage{graphicx}
\usepackage{epstopdf}
\usepackage{xcolor}
\usepackage[dvipsnames]{xcolor}
\usepackage{ulem}

\usepackage{setspace}

\begin{document}

\title{Static Friction of Liquid Marbles}

\author{Yui Takai}
\thanks{These authors contributed equally to this work.}
\affiliation{The University of Tokyo, Department of Mechanical Engineering, Tokyo, Japan.}

\author{Kei Mukoyama}
\thanks{These authors contributed equally to this work.}
\affiliation{The University of Tokyo, Department of Mechanical Engineering, Tokyo, Japan.}

\author{Pritam Kumar Roy}
\affiliation{The University of Tokyo, Department of Mechanical Engineering, Tokyo, Japan.}
\affiliation{Central University of Rajasthan, Department of Physics, Rajasthan, India}

\author{Guillaume Lagubeau}
\affiliation{Université de Lille, CNRS, UMR 8520-IEMN, Lille, France.}

\author{David Quéré}
\affiliation{ESPCI Paris, Physique et Mécanique des Milieux Hétérogènes, UMR 7636 du CNRS, Paris, France.}

\author{Samuel Poincloux}\email{poincloux@phys.aoyama.ac.jp}
\affiliation{Department of Physical Sciences, Aoyama Gakuin University, Sagamihara, Japan.}

\author{Timothée Mouterde}\email{mouterde@g.ecc.u-tokyo.ac.jp}
\affiliation{The University of Tokyo, Department of Mechanical Engineering, Tokyo, Japan.}

\begin{abstract}
Liquid marbles, droplets coated with a granular layer, are highly mobile as particles prevent capillary adhesion to the substrate. Yet their coating creates a static rolling friction, which we measure and model. Motion requires shear within the shell so that it is governed mainly by the grain density. This density controls yielding via a logistic function emerging as the particle network approaches percolation and increasing rapidly at close packing. More broadly, our system offers a simple platform for probing granular-raft mechanics and measuring their effective surface tension.

\end{abstract}

\maketitle

Manipulating small liquid volumes is challenging owing to capillary forces that pin droplets below the millimeter scale. A common approach to prevent capillary pinning is to form non-wetting states by separating the liquid from the solid via a gas layer. This layer can be formed actively by air entrainment from a fast-moving surface~\cite{lhuissier2013levitation,gauthier2016aerodynamic} or, in the Leidenfrost effect, by evaporation~\cite{biance2003leidenfrost}. On superhydrophobic surfaces, air is instead passively stabilized by hydrophobic micro/nanostructures~\cite{shibuichi1996super,blossey2003self}. Alternatively, structures can be embedded at the liquid surface by coating the droplet with particles, to form liquid marbles~\cite{aussillous2001liquid}, whose non-wetting state is substrate-independent. Particle properties can further be tailored to functionalize droplets, enabling manipulation by external stimuli such as light~\cite{paven2016light} or magnetic fields~\cite{zhao2010magnetic}, as well as the handling of hot liquids~\cite{roy2025hot}. This makes liquid marbles controllable and possibly usable as microreactors~\cite{sheng2015silica, oliveira2017potential,vadivelu2017liquid, tian2010liquid}, in soft robotics~\cite{jeon2025particle}, or materials design~\cite{rong2019liquid}.
More broadly, liquid--grain interactions are central to erosion processes~\cite{trottet2025sandball}.

Both marbles and droplets on superhydrophobic surfaces exhibit very low static friction, and even slight slopes can initiate their motion. For droplets, this friction arises from contact angle hysteresis and capillary forces~\cite{pockels1914randwinkel,furmidge1962studies}. In contrast, liquid marble static friction resembles solid-solid friction for moderate volumes $10$--$50~\mu\mathrm{L}$~\cite{jin2023marbles}, yet its physical origin remains largely unexplored. This paper examines how liquid marble's static friction arises from coupling between rolling deformation of aspherical droplets and their granular raft coating.

\begin{figure*}[]
	\centering
	\includegraphics[scale=0.29]{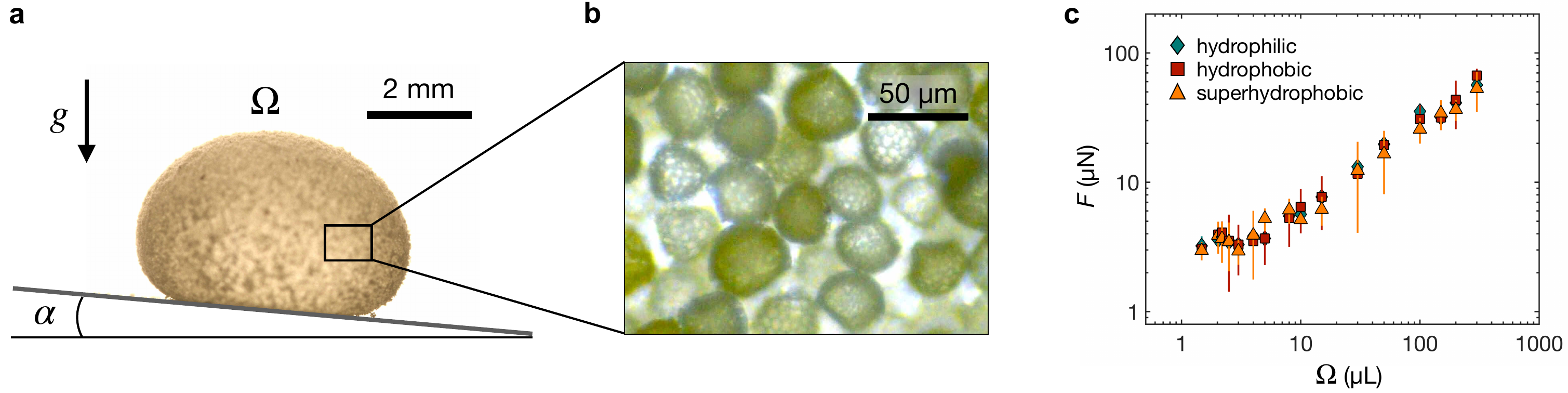}
	\caption{(a) A water marble with volume $\Omega$ is deposited on a silicon wafer, which is slowly tilted until the marble departs at angle $\alpha$. The photograph shows a marble with $\Omega=50~\mu\mathrm{L}$.
    (b) Microscope image of the hydrophobic lycopodium particles, of diameter $32\pm3~\mu\mathrm{m}$, used to form the granular shell.
    (c) Static friction $F$ of liquid marbles versus volume $\Omega$ on log--log scales for hydrophilic (blue diamonds), hydrophobic (red squares), and superhydrophobic substrates (orange triangles).}
	\label{F1}
\end{figure*}

Liquid marbles are prepared by coating water drops of surface tension $\gamma=72~\mathrm{mN/m}$, density $\rho=1000~\mathrm{kg/m^3}$, and volume $\Omega$ (Fig.~\ref{F1}a) with lycopodium particles (Fig.~\ref{F1}b). These porous hydrophobic particles have mean diameter $d=32~\mu$m, with standard deviation $3~\mu$m, and cover a surface fraction $\phi$ of the liquid interface. This preparation yields a reproducible initial surface fraction, $\phi_\mathrm{o}=0.83\pm0.01$ (End Matter, Appendices A, B). The particles lower the effective surface tension $\gamma_\mathrm{eff}(\phi)$~\cite{aussillous2006properties,mchale2011liquid}, which is estimated from the marble shape as $\gamma_\mathrm{eff}(\phi_\mathrm{o})=53\pm1~\mathrm{mN/m}$.
The marbles are then placed on a horizontal wafer, gradually tilted until the departure angle $\alpha$, at which the marbles start moving (Fig.~\ref{F1}a). Measured with a digital goniometer (Monotaro, Japan), $\alpha$ corresponds to the slope at which the tangential weight $F=\rho\Omega g\sin\alpha\simeq\rho\Omega g\alpha$ overcomes static friction. The lycopodium mass is neglected, as it contributes less than $3\%$ of the total mass.

\begin{figure}[]
	\centering
	\includegraphics[scale=0.33]{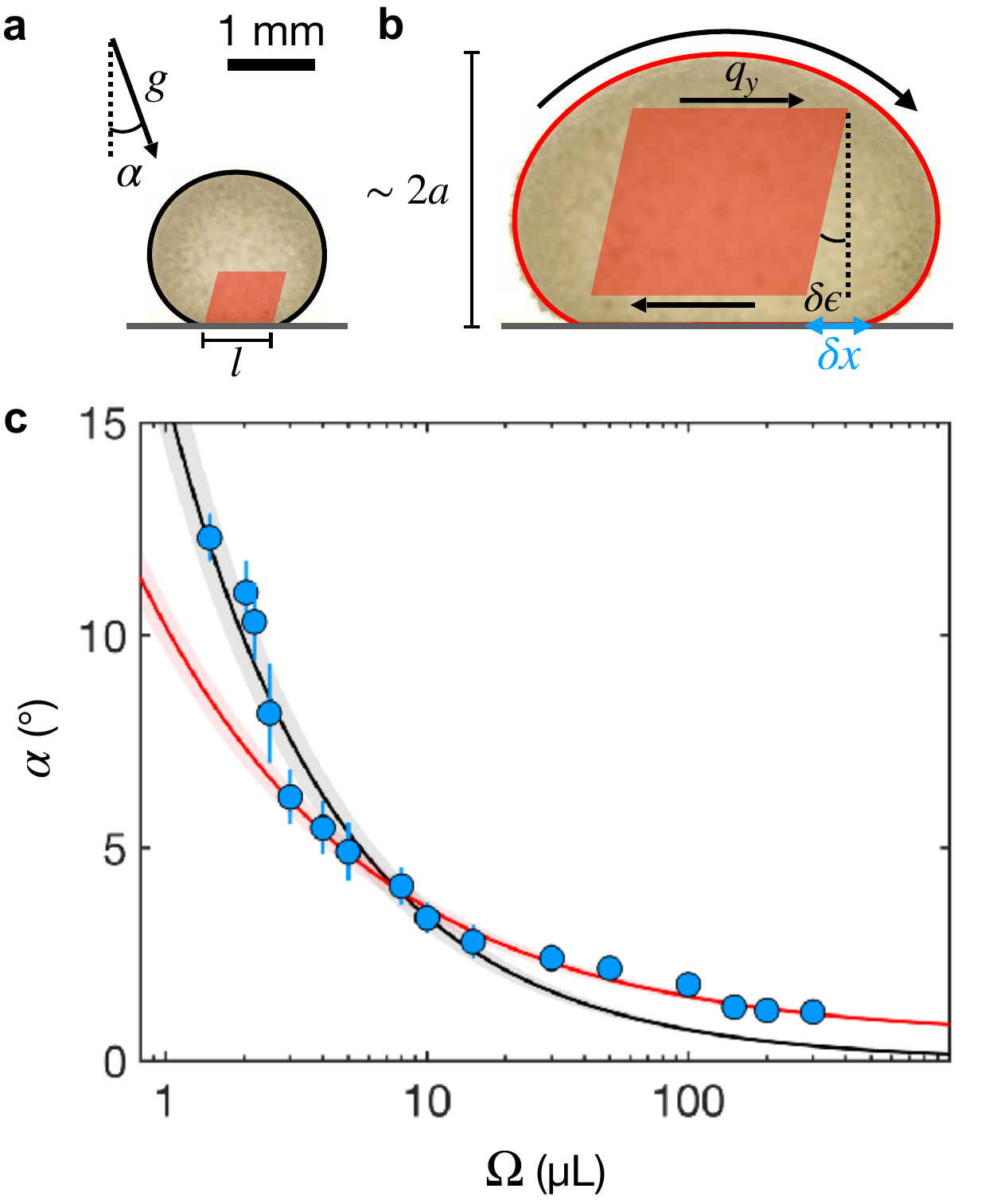}
	\caption{ (a-b) Sketches of liquid marbles rolling on a surface tilted by $\alpha$. The red regions denote the sheared part of the shell for marbles (a) smaller and (b) larger  than the capillary length. (c) Departure angle $\alpha$ versus liquid-marble volume $\Omega$ on a lin--log scale. Data are averages over the three substrates shown in Fig.~\ref{F1}c, since static friction is substrate-independent. The red line is Eq. \ref{eq:model_large} with $f(\phi_\mathrm{o}) = 0.26$, $a(\phi_\mathrm{o})= 2.3$~mm and $\mu$~=~0.053~$\pm$~0.003; the black line is Eq.~\ref{eq:model_small} adjusted by a factor of 0.69. 
    Solid lines indicate the fitted ranges, and shaded areas show the 95\% confidence intervals. }
		\label{F2}
\end{figure}

We first examine the effects of substrate properties and marble volume on static friction. Hydrophilic, hydrophobic, and silica-nanobead-coated superhydrophobic silicon wafers are prepared (End Matter, Appendix C). We vary the marble initial volume from $1.5$ to $300~\mu\mathrm{L}$ and measure the static friction $F$ on each substrate (Fig.~\ref{F1}c).
Three key observations emerge. (i) Static friction is substrate-independent despite differences in wetting and roughness and thus solid–solid friction. This rules out liquid– or particle–substrate dissipation. (ii) The marbles do not slide but initially roll (Supplemental Video 1, \cite{Supplemental}), so $F$ is a static rolling friction.
(iii) Below $7\pm2~\mu$L, static friction is nearly volume-independent, of order a few $\mu$N; above, it increases with volume up to $60~\mu$N at $300~\mu$L, more weakly than the linear scaling expected for Amontons--Coulomb sliding friction.
In both cases, the friction is a small fraction of the marbles' weight---typically less than 10$\%$---a hallmark of liquid marbles. 
The presence of two regimes suggests that the shape of the marbles influences their friction. This shape transition is governed by the capillary length $a(\phi) = \sqrt{\gamma_{\mathrm{eff}}(\phi)/\rho g}$: small marbles ($\Omega \ll a^3$) are nearly spherical (Fig.~\ref{F2}a), while larger ones are flattened by gravity (Fig.~\ref{F2}b). For $\phi = \phi_\mathrm{o}$, we find $a^3(\phi_\mathrm{o}) \approx 12$~$\mu\mathrm{L}$, consistent with the observed transition volume.

These observations suggest that static friction arises from the granular shell’s resistance to rolling, which we now model.
In classic rolling friction~\cite{reynolds1876vi}, resistance arises from viscoelastic bulk dissipation near the flattened contact region~\cite{brilliantov1998rolling}.
Liquid marbles also are never perfectly spherical, even below the capillary length, because of gravity.
Their rolling resistance likewise stems from asphericity, but deformation and dissipation occur in the granular shell rather than the liquid bulk: rolling without sliding requires geometrical rearrangements of the shell, accommodated by shear in the granular layer.

Dissipation by in-plane shear yielding of the granular shell, modeled as a granular raft~\cite{cicuta2009granular,protiere2023particle} wrapped around the droplet, thus governs the onset of rolling.
We determine the departure angle $\alpha$ from an energy argument. 
Rolling over a distance $\delta x$ begins when the gravitational energy gain $\delta E_g\simeq\rho g\Omega\alpha \ \delta x$ exceeds the shear dissipation $\delta E_d$ in the granular shell. 
Since rolling starts under quasi-static substrate tilting, viscous dissipation in the liquid is negligible (End Matter, Appendix D).
We assume that rolling by $\delta x$ imposes a homogeneous shear strain $\delta \epsilon$ over an area $S$ of the granular shell (Fig.~\ref{F2}a, b). To capture the onset of rolling, we model the shell rheology as purely plastic and quasi-static. The energy dissipated in this shell of volume $Sd$, then writes $\delta E_d\sim S d q_y \delta\epsilon$, with $q_y$ the yield stress of the granular shell~\cite{kruyt2006shear}.
This estimate neglects the elastic response of the granular assembly by assuming that the imposed shear is fully dissipated. We use a cohesionless Mohr--Coulomb criterion, $q_y=\mu p(\phi)$, where $\mu$ is a macroscopic friction coefficient~\cite{andreotti2013granular} and $p(\phi)$ is the mean normal stress in the shell, taken homogeneous over the marble.
This stress reduces the marble's effective surface tension, $\gamma_{\mathrm{eff}} = \gamma-p(\phi)d$~\cite{cicuta2009granular,lagubeau2014armoring}.
Out-of-plane deformations and dynamic rolling would require additional ingredients.
We rewrite the pressure term as $p(\phi)= \gamma f(\phi)/d$, introducing a dimensionless function $f(\phi)$ that captures the surface-fraction dependence. 
Microscopically, $f(\phi)$ represents the dimensionless average normal component of intergrain contact forces.
The effective surface tension is then  $\gamma_{\mathrm{eff}} = \gamma(1-f(\phi))$, so $f$ also quantifies how $\gamma_\mathrm{eff}$ varies with $\phi$.
 Substituting this into the energy balance $\delta E_g\sim \delta E_d$, we obtain the departure angle:

\begin{equation}
     \alpha \approx \frac{S a^2}{\Omega}\frac{\delta\epsilon}{\delta x}\mu \frac{f(\phi)}{1-f(\phi)}
    \label{eq:model_general}
\end{equation}

Here, the sheared shell area $S$ and the shear strain $\delta \epsilon$ involved in dissipation are primarily set by the marble's aspheric deformation, analogous to shear in rolling non-wetting droplets~\cite{mahadevan1999rolling}.
Gravity flattens the marbles, forming a contact disk of diameter $l$ and setting the sheared region's height (Fig. \ref{F2}a, b).
We consider the large-marble regime of Fig.~\ref{F1}c, where gravity flattens the marbles to a height $\simeq2a$. 
Approximating their shape as a cylinder of base diameter $l$ and height $2a$ gives $\Omega\simeq\pi l^2a/2$.
The flat base of the large marbles being more extended than their height, we take the sheared area to be the entire cylindrical surface, $S\approx \pi l^2/2+2\pi l a$, and the shear length as the marbles' height $2a$, yielding $\delta\epsilon\approx\delta x/2a$ (Fig.~\ref{F2}b).

Combining these with Eq.~\ref{eq:model_general} gives the departure angle:
\begin{equation}
    \alpha\approx  \frac{1}{2}\left({1+\sqrt{\frac{8\pi a^3}{\Omega}}}\right)\mu \frac{f(\phi)}{1-f(\phi)} \, ,
    \label{eq:model_large}
\end{equation}

To compare the model with experiments, we plot in Fig.~\ref{F2}c the departure angle averaged over the three substrates shown in Fig.~\ref{F1}c, since static friction is substrate-independent. Fitting Eq.~\ref{eq:model_large} in the large-marble regime of Fig.~\ref{F1}c, $\Omega>7~\mu$L, with $f(\phi_\mathrm{o})=0.26$ and $a(\phi_\mathrm{o})\simeq2.3~\mathrm{mm}$, i.e. $\gamma_\mathrm{eff}(\phi_\mathrm{o})\simeq53~\mathrm{mN/m}$, gives $\mu=0.053 \pm 0.003$ (red line in Fig.~\ref{F2}c). The model agrees with the data over two decades in volume, from $3$ to $300~\mu$L, with deviations only for $\Omega<3~\mu$L.

To capture this deviation, we adapt the model to small, quasi-spherical marbles of radius $R$, with $\Omega\simeq4\pi R^3/3$. Balancing the weight, $\rho gR^3$, with the Laplace pressure integrated over the contact area, $\gamma_{\mathrm{eff}}l^2/R$, gives $l\sim R^2/a$~\cite{mahadevan1999rolling}. Most of the marble rotates as a rigid body, so shear is confined near the contact over a distance $l$ (Fig.~\ref{F2}a). Thus $S\simeq2\pi Rl$ and $\delta\epsilon\simeq\delta x/l$, which with Eq.~\ref{eq:model_general} gives for small marbles:
\begin{equation}
    \alpha \approx \left({6\pi^2}\right)^{1/3} \left( \frac{a^3}{\Omega}\right) ^{2/3} \mu \frac{f(\phi)}{1-f(\phi)}
    \label{eq:model_small}
\end{equation}
Using $\mu$ obtained for large marbles, Eq.~\ref{eq:model_small} predicts the small-$\Omega$ data with a prefactor of $0.69$ (black line, Fig.~\ref{F2}c), close to unity and accounting for geometry-dependent approximations. The small-marble model remains valid up to $\sim15~\mu$L, within experimental uncertainty.
These results call for several remarks. (i) For $3\lesssim\Omega\lesssim15~\mu$L, close to $a^3\simeq12~\mu$L, both models describe the data within error bars, corresponding to the shape transition from spherical to flattened marbles. (ii) The model predicts three successive regimes as $\Omega$ increases: $\alpha\sim\Omega^{-2/3}$ at small volumes, a crossover with $\alpha\sim\Omega^{-1/2}$, and a plateau at large volumes.
(iii) The small fitted value of the macroscopic friction coefficient reflects that granular rafts at $\phi_\mathrm{o}$ are fragile assemblies that yield easily under shear.

\begin{figure*}[]
	\centering
	\includegraphics[scale=0.33]{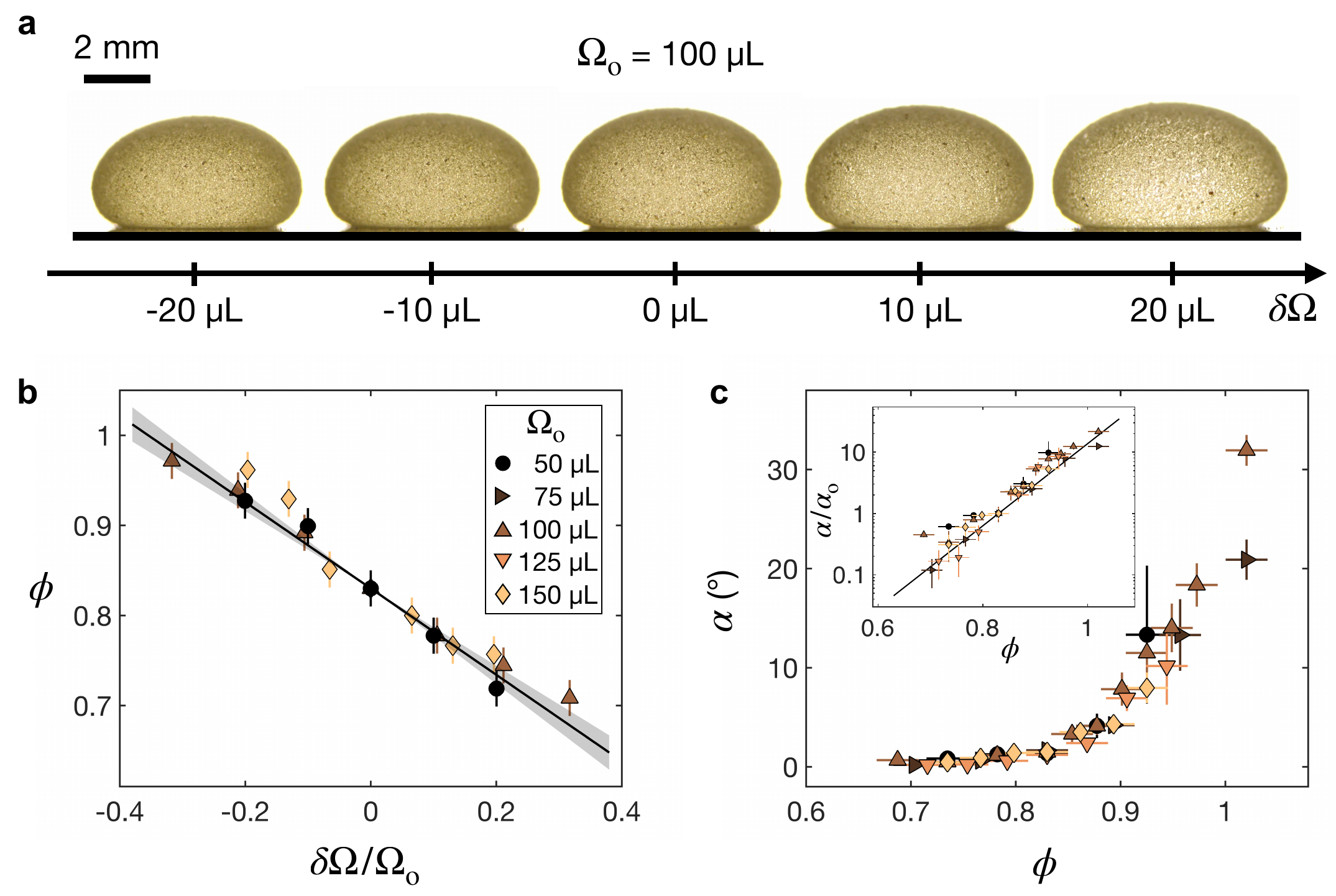}
	\caption{ (a) Photographs of lycopodium marbles with initial volume $\Omega_\mathrm{o}=100~\mu\mathrm{L}$, from which liquid is removed ($\delta\Omega<0$) or added ($\delta\Omega>0$) at fixed particle number. Here $\delta\Omega$ ranges from $-20$ to $20~\mu\mathrm{L}$; the central image corresponds to the standard preparation.
    (b) Particle surface fraction $\phi$ versus relative volume change $\delta\Omega/\Omega_\mathrm{o}$ for $\Omega_\mathrm{o}=50$, 75, 100, 125, and $150~\mu\mathrm{L}$. Surface areas and volumes are obtained by image analysis assuming axisymmetry. The black line is the fit $\phi=\phi_\mathrm{o}(1-0.58\delta\Omega/\Omega_\mathrm{o})$, with $\phi_\mathrm{o}=0.83$; the shaded area is the 95\% confidence interval.
    (c) Departure angle $\alpha$ versus particle surface fraction $\phi$ for $\Omega_\mathrm{o}=50$--$150~\mu\mathrm{L}$. Inset: $\alpha/\alpha_\mathrm{o}$, with $\alpha_\mathrm{o}=\alpha(\phi_\mathrm{o})$, plotted versus $\phi$ on a log-linear scale. The black line is the best exponential fit, $\exp[\beta(\phi-\phi_\mathrm{o})]$, with $\beta=15.4\pm1.7$.}
	\label{F3}
\end{figure*}

Our model also predicts that static friction depends on the intergrain normal force through $f(\phi)$. 
This function increases non-linearly with the particle surface fraction $\phi$ as observed in rafts and marbles~\cite{cicuta2009granular,varshney2012amorphous,lagubeau2014armoring}, but also more generally in 2D rheology of particles trapped at interfaces~\cite{cicuta2003shearing, madivala2009self,kato2023surface}. 
We test this hypothesis with a second experiment. Marbles are prepared with initial volume $\Omega_\mathrm{o}$, surface area $S_\mathrm{o}$, and surface fraction $\phi_\mathrm{o}$. $S$ is then varied by adding ($\delta\Omega>0$) or removing ($\delta\Omega<0$) water with a microsyringe (SGE Analytical Science). Since the number of particles, $\sim\phi_\mathrm{o}S_\mathrm{o}/d^2$, is fixed, the new surface fraction is $\phi=(S_\mathrm{o}/S)\phi_\mathrm{o}$.
This method is suitable only for marbles much larger than the tip of the microsyringe ($\Omega\geq50\,\mu\mathrm{L}$).
Figure~\ref{F3}a shows deflated and inflated $100~\mu\mathrm{L}$ marbles. From these images, we measure the total surface area $S$ using the Pappus--Guldinus theorem, assuming axisymmetry (End Matter, Appendix E). Figure~\ref{F3}b shows the surface fraction $\phi$ versus the relative volume change $\delta\Omega/\Omega_\mathrm{o}$ for marbles with $\Omega_\mathrm{o}=50$, 75, 100, 125, and $150~\mu\mathrm{L}$. The surface fraction decreases linearly with $\delta\Omega/\Omega_\mathrm{o}$ and is well fitted by $\phi=\phi_\mathrm{o}(1-0.58\delta\Omega/\Omega_\mathrm{o})$. We use this relation to estimate $\phi$ for all marbles.

For the resulting marbles, with volume $\Omega$, surface area $S$, and surface fraction $\phi$, we measure the departure angle $\alpha$ as a function of $\phi$ for initial volumes $\Omega_\mathrm{o}=50$--$150~\mu\mathrm{L}$.
Just before rolling starts a front--rear curvature difference $\Delta\mathcal{C}$ develops despite the absence of classical contact-angle hysteresis (Supplemental Video 2, \cite{Supplemental}). This deformation becomes more pronounced as $\phi$ approaches 1 (Fig.~\ref{FSI1}). 
In short, the curvature asymmetry induces a Laplace pressure difference $\gamma_\mathrm{eff}\Delta\mathcal{C}$ that balances the hydrostatic pressure variation $\rho g l\alpha$. At fixed tilt, increasing $\phi$ lowers $\gamma_\mathrm{eff}$, requiring larger deformations to maintain this balance (End Matter, Appendix F).
The variation of the departure angle with surface fraction $\alpha(\phi)$, is plotted in Fig.~\ref{F3}c. 
The departure angle $\alpha$, and thus static friction, increases exponentially with $\phi$. Data for all initial volumes collapse onto $\alpha / \alpha_\mathrm{o} = \exp[\beta(\phi - \phi_\mathrm{o})]$, with $\beta = 15.4 \pm 1.7$ (inset, black line), where $\alpha_\mathrm{o}$ is the sliding angle at $\phi_\mathrm{o}$.
The departure angle, which corresponds to the friction-to-weight ratio, abruptly increases from less than 0.8$\%$ for inflated marbles with $\phi \sim 0.7$ to more than 50$\%$ for deflated marbles with $\phi$ close to 1.
This effect dominates over the weak variations due to the initial marble volume in the large marbles regime (Eq.~\ref{eq:model_large}). This observation also reveals that a non-wetting condition is not sufficient to ensure high mobility; a relatively low surface fraction is also necessary.

\begin{figure}[]
	\centering
	\includegraphics[scale=0.36]{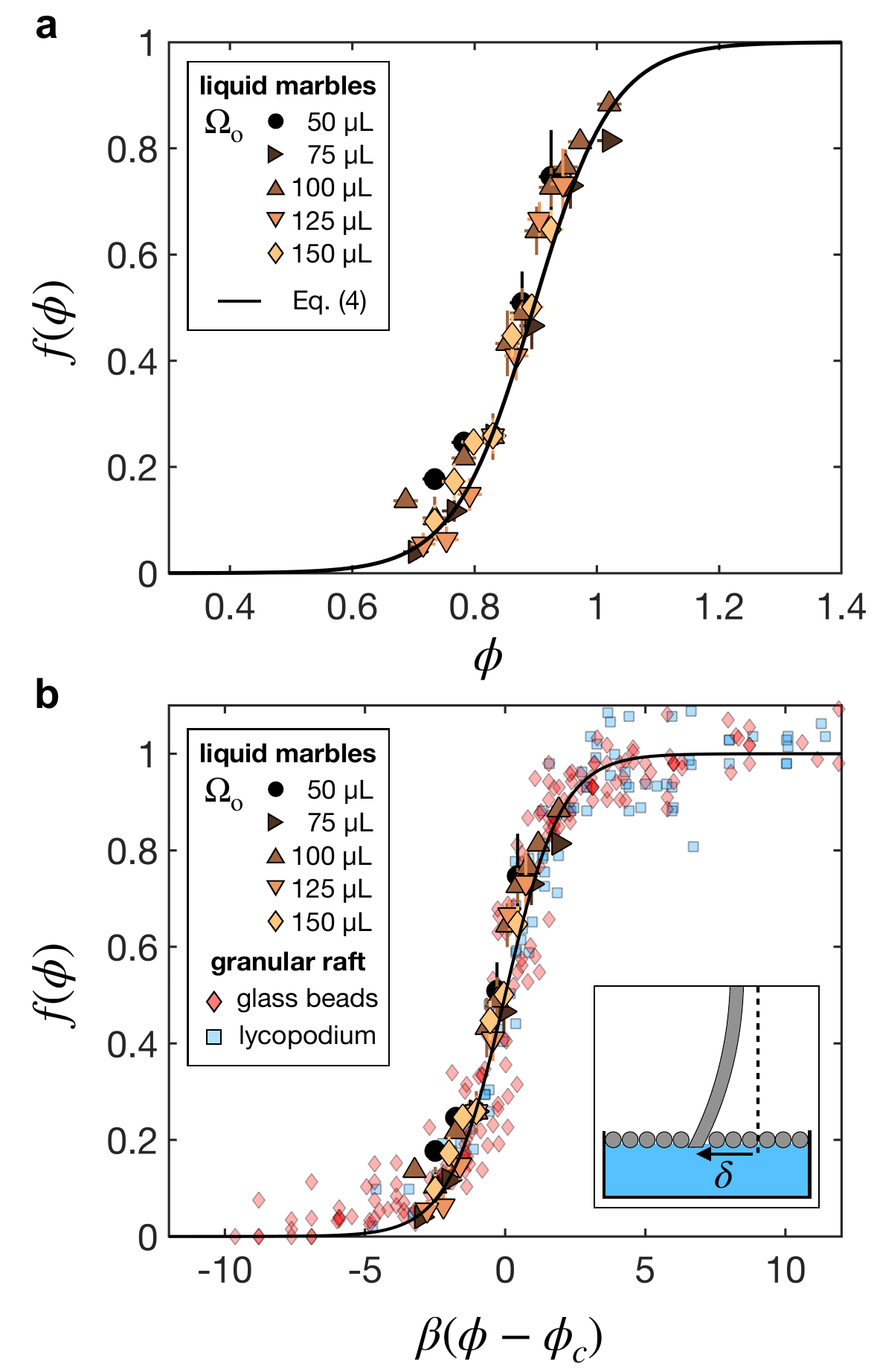}
	\caption{
(a) $f(\phi)$ versus $\phi$, obtained from the departure angles of inflated and deflated liquid marbles with initial volumes $\Omega_\mathrm{o}=50$--$150~\mu\mathrm{L}$, color-coded from black to beige. Values are obtained using Eq.~\ref{eq:model_large} in the asymptotic large-volume regime. The black line is Eq.~\ref{eq:f_th}. (b) $f(\phi)$ versus $\beta(\phi-\phi_c)$ for liquid marbles, same data as in (a), and force measurements in two-dimensional granular rafts made of glass beads (red diamonds) or lycopodium particles (blue squares). The black line is the logistic function of Eq.~\ref{eq:f_th}. Inset: schematic of the granular-raft yielding experiment, where the force is obtained from the deflection $\delta$ of a calibrated fiber.}
	\label{F4}
\end{figure}

From this experimental result and using Eq.~\ref{eq:model_large} in the asymptotic limit $\Omega/a^3\gg 1$, we obtain an analytical estimate for $f(\phi)$ in the form of a logistic function:
\begin{equation}
    f(\phi) = \frac{1}{1+e^{-\beta(\phi-\phi_c)}} ,
    \label{eq:f_th}
\end{equation}
with $\phi_c = \phi_\mathrm{o} - (1/\beta)\ln\left( \gamma/\gamma_\mathrm{eff}(\phi_\mathrm{o})-1\right) $, that is $\phi_c \approx$ 0.90~$\pm$~0.01. This approximation is equivalent to assuming that dissipation occurs mainly at the top and bottom of the marbles, yet the full numerical solution yields curves very close to the logistic form (End Matter, Appendix G and Fig.~\ref{F_noapprox}).
We also note that $\phi_\mathrm{c}$ is independent of the choice of $\phi_\mathrm{o}$, despite its appearance in the expression.
Fig.~\ref{F4}a shows $f(\phi)$ estimated directly from the departure angles using Eq.~\ref{eq:model_large}, together with the prediction of Eq.~\ref{eq:f_th}, both showing good agreement.
We then interpret geometrically the logistic form of $f(\phi)$ and its characteristic surface densities.
Yielding increases progressively from $\phi\simeq0.65$, close to the 2D percolation threshold for monodisperse disks, $0.676$~\cite{quintanilla2000efficient}. Above this threshold, a connected particle network forms, so that added particles increase the number of contacts across the shell and amplify the contact pressure.
The inflection point occurs at $\phi_c\simeq0.90$, close to the close-packing fraction of 2D disks, $\pi/\sqrt{12}\simeq0.907$~\cite{toth2014regular}.
This suggests that saturation arises when, beyond close packing and up to surface fractions exceeding one, the interface deforms and buckles, redirecting added-particle forces out of plane rather than increasing in-plane normal forces between grains $f(\phi)$. 
Remarkably, the expression of $f(\phi)$ in Eq.~\ref{eq:f_th}, obtained assuming a continuous and homogeneous surface density and pressure field, remains valid even at low $\phi$ where gaps may appear, and at high $\phi$ where wrinkles develop. 
This saturation by wrinkling is observed both in marbles~\cite{lagubeau2014armoring} and rafts~\cite{vella2004elasticity,jambon2017wrinkles,protiere2023particle,prakash2024buckling}.

To test the generality of the logistic prediction beyond liquid marbles, we turn to two-dimensional granular rafts, another type of interfacial granular assembly. For rafts, the effective confining pressure can be probed directly by measuring the force required to yield the granular in an obstacle-dragging experiment~\cite{dollet2005two, kolb2013rigid, seguin2016local}.
Rafts are prepared by depositing a mass $m_g$ of hydrophobic glass beads (water contact angle 110°) or lycopodium, of diameter $d$ and density $\rho_g$, onto a water bath in a Langmuir–Blodgett trough (End Matter, Appendix H). Adjusting the bath area $S_t$ controls the monolayer surface fraction, $\phi=3m_g/(2\rho_g dS_t)$.
To measure intergrain forces, we insert vertically the core of a glass optical fiber into the raft. The fiber, of diameter $125~\mu$m and length $3$--$6$~cm depending on the experiment, is clamped at one end while the trough is translated horizontally at $1~\mathrm{mm/s}$, slow enough to avoid dynamic effects. As the raft resists the fiber motion, the fiber bends, and the yielding force $F_y$ is obtained from its deflection $\delta$ (Fig.~\ref{F4}b, inset) using $F_y=k\delta$. The effective stiffness $k$ is calibrated from the static deflection of the horizontally clamped fiber under its own weight, giving $k=10$--$60~\mathrm{mN/m}$ for long to short fibers, in agreement with the cantilever-beam equation (End Matter, Appendix I).
In steady state, the fiber locally yields the granular raft, and we expect the exerted force to scale as $F_y \sim q_y$, and thus with $F_y\sim f(\phi)$.
Since $f(\phi)$ is expected to saturate to unity at large surface fractions, we normalize $F_y(\phi)$ by its asymptotic value $F_y^{\max}$. This provides an independent probe of the functional form $f(\phi)=F_y(\phi)/F_y^{\max}$ and, consequently, of the interface effective surface tension.

The measurements reveal that $f(\phi)$ for rafts also follows a logistic function (Fig. \ref{f_data}). For each curve, we extract $\beta$ and $\phi_c$ by fitting $f(\phi)$ with Eq.~\ref{eq:f_th} (Table~\ref{table_fit}).
In Fig.~\ref{F4}b, we plot $f(\phi)=1-\gamma_\mathrm{eff}/\gamma$ versus $\beta(\phi-\phi_c)$. All data from marbles and rafts collapse onto the logistic prediction, spanning the full range $0\leq\gamma_\mathrm{eff}/\gamma\leq1$. The agreement between marble- and raft-based measurements confirms that the static rolling friction of marbles is governed by yielding of their granular shell, which behaves as a two-dimensional raft. Moreover, the logistic form of $f(\phi)$ appears independent of particle material and measurement method, suggesting an intrinsic structural property of granular rafts. While the shape seems universal, the steepness $\beta$ and inflection density $\phi_c$ vary significantly between marbles and rafts, but also between experimental runs for the rafts. 
We attribute this variability to the preparation protocol: marble shells are reproducibly prepared in a dense state, $\phi\simeq0.8$, by repeated rolling over a granular bed, whereas rafts are prepared at low density, $\phi\simeq0.5$, and then quasi-statically compressed without mechanical annealing. Their microstructure therefore remains protocol- and history-dependent~\cite{planchette2025unjamming}, accounting for the run-to-run variability in $\beta$ and $\phi_c$.

In conclusion, our findings show that the particle shell responsible for the high mobility of liquid marbles also gives rise to their static rolling friction. Rolling requires shear-induced yielding of the granular shell, governed mainly by particle surface density, with marble shape weakly correcting the shear profile and yielding region.
Despite being nearly spherical, smaller marbles require larger tilt angles to move under gravity than larger, gravity-flattened marbles. This contrasts with their dynamics, where smaller marbles roll faster~\cite{mahadevan1999rolling,aussillous2001liquid}. The particle surface density $\phi$ controls yielding through a logistic function that captures the shell rheology: yielding begins near the percolation threshold, while the inflection point coincides with the close-packing limit of the granular layer. The physical mechanism setting the steepness parameter $\beta$, however, remains an open question.
Finally, our results underscore the critical role of particle surface fraction in maintaining low static friction, especially when evaporation increases shell density over time. Liquid marbles thus provide a platform to explore the interplay between two-dimensional interfacial rheology~\cite{fuller2012complex,ji2020interfacial} and the macroscopic mechanics of such interfaces embedded in three-dimensional geometries.
\\

\begin{acknowledgments}
\paragraph*{Acknowledgements}
We thank Dominic Vella for fruitful discussions. 
TM acknowledges the financial support provided by the Japan Society for the Promotion of Science (JSPS) - Grant-in-Aid for Scientific Research (B), 24K01341. SP acknowledges the financial support provided by JSPS KAKENHI Grant-in-Aid for Research Activity Start-up 24K22937. TM and PKR gratefully acknowledge the financial support provided by the JSPS Grant-in-Aid for JSPS Fellows 23KF0106. YT is grateful for the financial support from JST SPRING (Grant Number JPMJSP2108) and the Leadership Development Program for Ph.D. students at the University of Tokyo, School of Engineering.
\end{acknowledgments}

\paragraph*{Data Availability}
The data that support the findings of this article is openly available in Zenodo
\cite{dataset}.




\appendix

\setcounter{figure}{0}
\renewcommand{\thefigure}{S\arabic{figure}}

\section{End Matter}

\textit{Appendix A: Liquid marble preparation ---} Liquid marbles are prepared by depositing a water volume $\Omega$ with a micropipette onto a bed of lycopodium particles (Sigma-Aldrich). The droplet is gently rolled until fully coated with particles. The resulting liquid marble is transferred to a small stainless-steel Petri dish with a superhydrophobic metal spoon, rolled a few times to remove excess particles, and placed on the silicon wafer for static-friction measurements.
\\

\textit{Appendix B: Experimental determination of the surface fraction ---} The surface fraction $\phi_\mathrm{o}$ is determined directly from side-view photographs. The liquid surface appears brighter than the particles, allowing the surface fraction to be estimated by intensity thresholding. We measured $\phi_\mathrm{o}$ for liquid marbles with volumes $\Omega = 2, 5, 7, 10, 30, 50, 100, 200,$ and $300~\mu\mathrm{L}$, using four marbles per volume. We obtained an average surface fraction of $83.0\%$, with a standard deviation of $1.25\%$.
\\

\textit{Appendix C: Substrate preparation ---} The hydrophilic substrate is a pristine silicon wafer with advancing and receding water contact angles of $23 \pm 2^\circ$ and $12 \pm 2^\circ$, respectively. The hydrophobic substrate is a silicon wafer whose surface is activated for $40$ s in a plasma cleaner and then coated by chemical vapor deposition of 1H,1H,2H,2H-perfluorodecyltrichlorosilane (Fluorochem Ltd.). The resulting advancing and receding water contact angles are $126 \pm 2^\circ$ and $99 \pm 2^\circ$, respectively. The superhydrophobic substrate is obtained by spraying a commercial coating (Glaco Mirror Coat Zero, Soft99) onto a silicon wafer. The substrates are dried vertically and, after solvent evaporation, heated on a hot plate at $200^\circ$C for $30$ min. This process is repeated three times. The resulting coating is a rough layer of hydrophobic silica nanobeads, with advancing and receding water contact angles of $161 \pm 2^\circ$ and $152 \pm 2^\circ$, respectively.
\\

\textit{Appendix D: Departure angle measurement method ---} The substrate is attached to a rotating stage controlled by a micrometer screw for fine adjustment. It is tilted at a typical angular speed $\dot\alpha \sim 2\times10^{-3}$ rad/s until the marble starts moving. The departure angle is then measured with a digital goniometer (Monotaro, Japan). To quantify possible dynamic lubrication effects between grains, we compare the characteristic viscous stress $\eta \dot\alpha$, where $\eta$ is the liquid dynamic viscosity, to the normal stress $\gamma f(\phi)/d$. This gives the viscous number~\cite{boyer2011unifying} $J = \eta \dot\alpha d/\gamma f(\phi)$. This number diverges below the percolation threshold as the normal force between grains vanishes, $f(\phi) \rightarrow 0$. However, over the range of surface fractions explored in our experiments, the minimum value of $f(\phi)$ is about $10^{-2}$. Even for this small value, we find $J \approx 10^{-7} \ll 1$, indicating that dynamic effects are negligible.
\\

\textit{Appendix E: Measurement of the marble surface area ---} To estimate the change in surface fraction induced by volume changes, we measure the surface area of liquid marbles from side-view images. Each image is binarized, and we identify the vertical symmetry axis dividing the marble into two equal halves. Assuming axisymmetry about this axis, we analyze one half of the profile. We detect its edge, which defines a planar curve $\mathcal{C}$. The marble surface is then modeled as the surface of revolution generated by rotating $\mathcal{C}$ about the symmetry axis. We compute the arc length $s$ of $\mathcal{C}$ and the distance $X$ of its centroid from the symmetry axis. The total surface area is then obtained from the Pappus--Guldinus theorem as $S = 2\pi s X$. This axisymmetric approximation may become less accurate as the surface fraction approaches unity and buckling occurs.
\\

\begin{figure}[]
	\centering
	\includegraphics[scale=0.33]{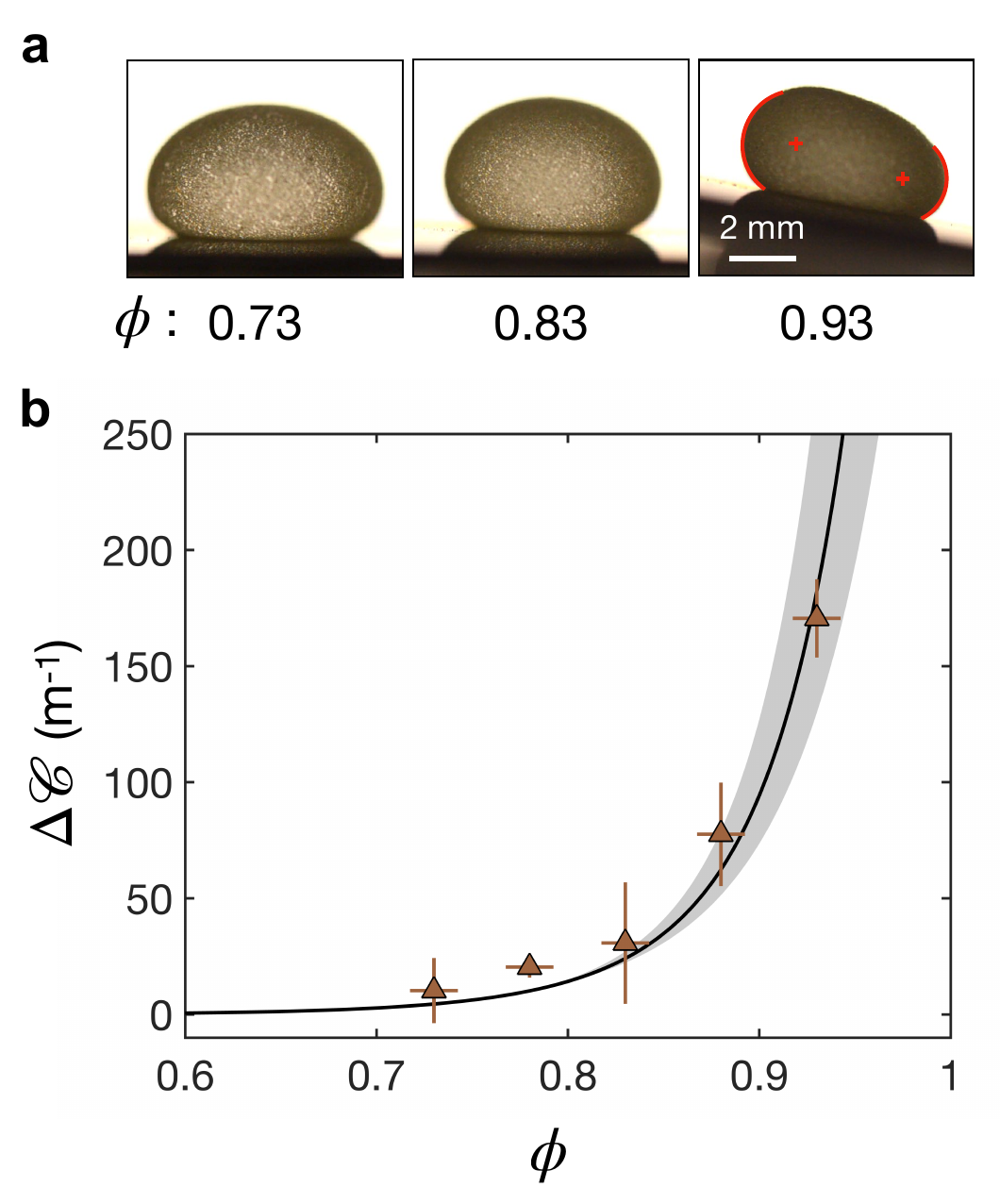}
	\caption{ (a) Images of liquid marbles just before they depart from the substrate with volume 100 $\mu$L and with $\phi$ varied from left to right from 0.73, 0.83 and 0.93. The images show that larger surface fractions lead to greater deformations. (b) Curvature difference $\Delta \mathcal{C}$ between the rear and front side of the liquid marbles as a function of the surface fraction $\phi$ for liquid marbles with initial volume 100 $\mu$L.}
		\label{FSI1}
\end{figure}

\textit{Appendix F: Liquid marble shape deformation ---} Figure~\ref{FSI1}a shows liquid marbles just before rolling for increasing surface fraction, from left to right: $\phi=0.73$, $0.83$, and $0.93$ (initial volume $\Omega_\mathrm{o}=100~\mu$L). The marbles exhibit a front--rear curvature difference $\Delta \mathcal{C}=\mathcal{C}_\mathrm{front}-\mathcal{C}_\mathrm{rear}$, measured from side-view images by fitting each interface with a circular arc of radius $\mathcal{C}^{-1}$. Figure~\ref{FSI1}b shows $\Delta \mathcal{C}$ as a function of the surface fraction $\phi$, revealing a pronounced increase in asymmetry at large $\phi$, which we now model. When the static liquid marble is tilted, the hydrostatic pressure difference between front and rear, $\rho g l \alpha$, is balanced by the Laplace pressure difference $\gamma_\mathrm{eff}\Delta \mathcal{C}$ associated with their curvature difference. Thus, $\Delta \mathcal{C}\approx \rho g l\alpha/\gamma_\mathrm{eff}$. Using volume conservation, $l\approx \sqrt{2\Omega/a\pi}$, and evaluating $\alpha$ at the departure angle, we obtain the curvature difference:
\begin{equation}
    \Delta \mathcal{C} \approx   \sqrt{\frac{\Omega}{2\pi a^5(\phi)}}\left({1+\sqrt{\frac{8\pi a^3(\phi)}{\Omega}}}\right)\mu \frac{f(\phi)}{1-f(\phi)} \, ,
    \label{eq:curvature}
\end{equation}
We plot this equation in Fig.~\ref{FSI1}b, using the empirical relation between volume and surface coverage $\phi = \phi_\mathrm{o} (1-0.58(\Omega-\Omega_\mathrm{o})/\Omega_\mathrm{o})$ and taking $\mu = 0.053$, the value obtained in the main text. Without any adjustable parameters, the model shows an excellent agreement with the measured curvature difference. This also reveals three mechanisms responsible for the large deformation observed before rolling at high particle fractions. (1) Increasing $\phi$ increases the departure angle and thus the hydrostatic pressure difference across the marble. This pressure difference must be balanced by a Laplace pressure difference, requiring a larger front--rear curvature difference. (2) Increasing $\phi$ decreases the effective surface tension, so a larger curvature difference is needed to produce the same Laplace pressure difference. (3) The reduced surface tension flattens the marble and increases its diameter $l$, further increasing the hydrostatic pressure difference and the required curvature difference.
\\

\begin{figure}[h]
	\centering
	\includegraphics[scale=0.33]{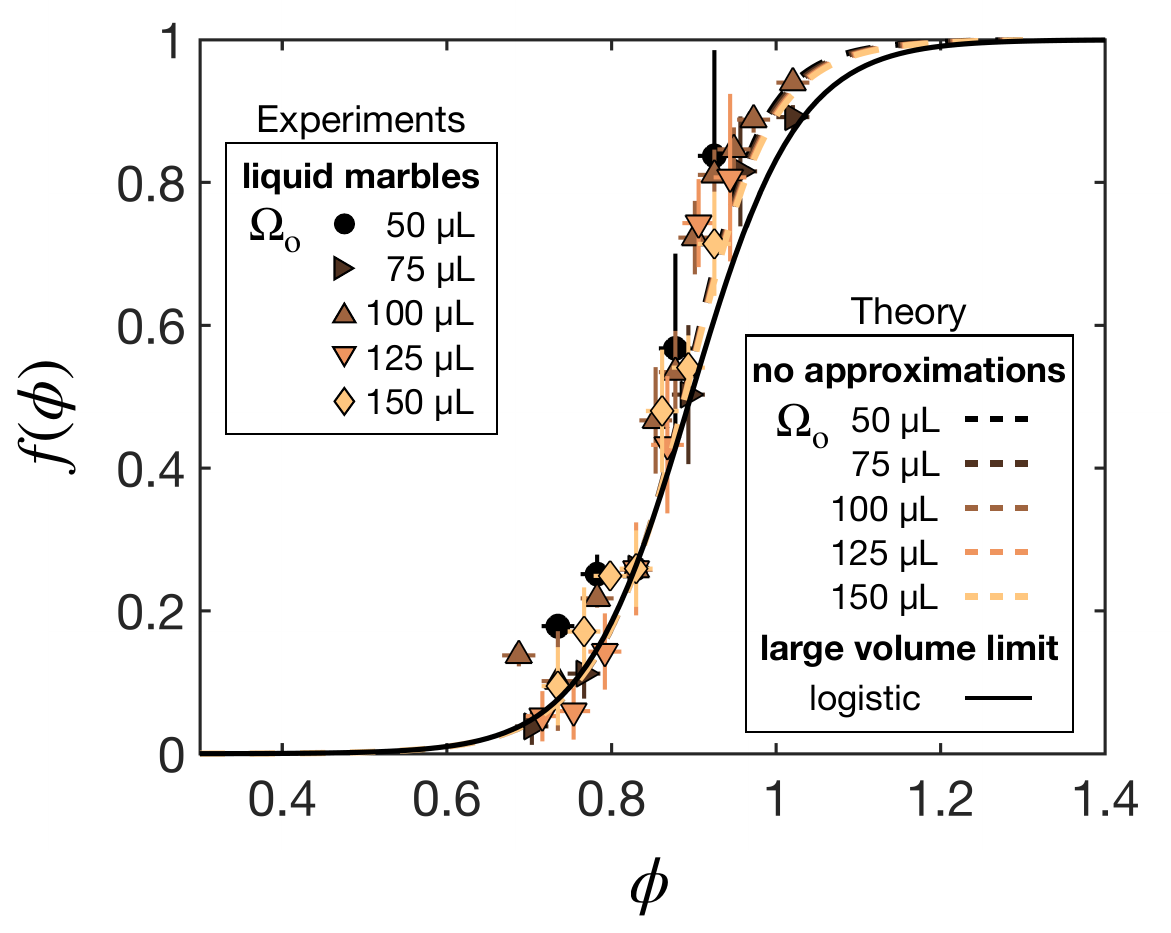}
	\caption{Comparison between the asymptotic limit of $f(\phi)$ given by Eq. \ref{eq:f_th} (black line) and the numerical solutions without approximations obtained by solving Eq. \ref{eq:f_no_approx} (dashed lines).
    }
	\label{F_noapprox}
\end{figure}

\textit{Appendix G: $f$ without approximations ---} The term $\sqrt{8\pi a^3/\Omega}$ neglected to derive Eq.~\ref{eq:f_th} can be as high as $\approx 2.5$ over the range of volumes and surface fractions explored. We estimate its impact on $f(\phi)$ by combining Eq.~\ref{eq:model_large} with the empirical relation $\phi = \phi_\mathrm{o}(1 - 0.58,\delta\Omega / \Omega_\mathrm{o})$. This gives:
\begin{equation}
    \frac{f(\phi_\mathrm{o})(1-f(\phi))}{f(\phi)(1-f(\phi_\mathrm{o}))} = \frac{\alpha_\mathrm{o}}{\alpha(\phi)} \frac{{1+\sqrt{\frac{8\pi a^3(\phi)}{\Omega}}}}{{1+\sqrt{\frac{8\pi a^3(\phi_\mathrm{o})}{\Omega_\mathrm{o}}}}},
    \label{eq:f_no_approx}
\end{equation}
with $f_\mathrm{o}=f(\phi_\mathrm{o})$ and $a_\mathrm{o}=a(\phi_\mathrm{o})$. Knowing $\alpha(\phi)/\alpha_\mathrm{o}$ for each pair of $(\Omega_\mathrm{o},\phi)$ values, Eq.~\ref{eq:f_no_approx} can be solved numerically for $f(\phi)$. For the experimental points, $\alpha(\phi)/\alpha_\mathrm{o}$ is taken from the measurements, whereas for the analytical curves we use the fitted empirical relation $\alpha(\phi)/\alpha_\mathrm{o}=\exp[\beta(\phi-\phi_\mathrm{o})]$. Without approximation, $f(\phi)$ also depends on the initial volume $\Omega_\mathrm{o}$, but this dependence is very weak, as shown in Fig.~\ref{F_noapprox}. The full numerical solution and the analytical approximation deviate slightly between $\phi\approx0.9$ and $\phi\approx1.2$, with the full solution remaining closer to the experimental data. However, since the deviation remains below $10\%$, we emphasize the simplified model for its compact formulation and easier interpretation.
\\

\begin{figure}[h]
	\centering
	\includegraphics[scale=0.33]{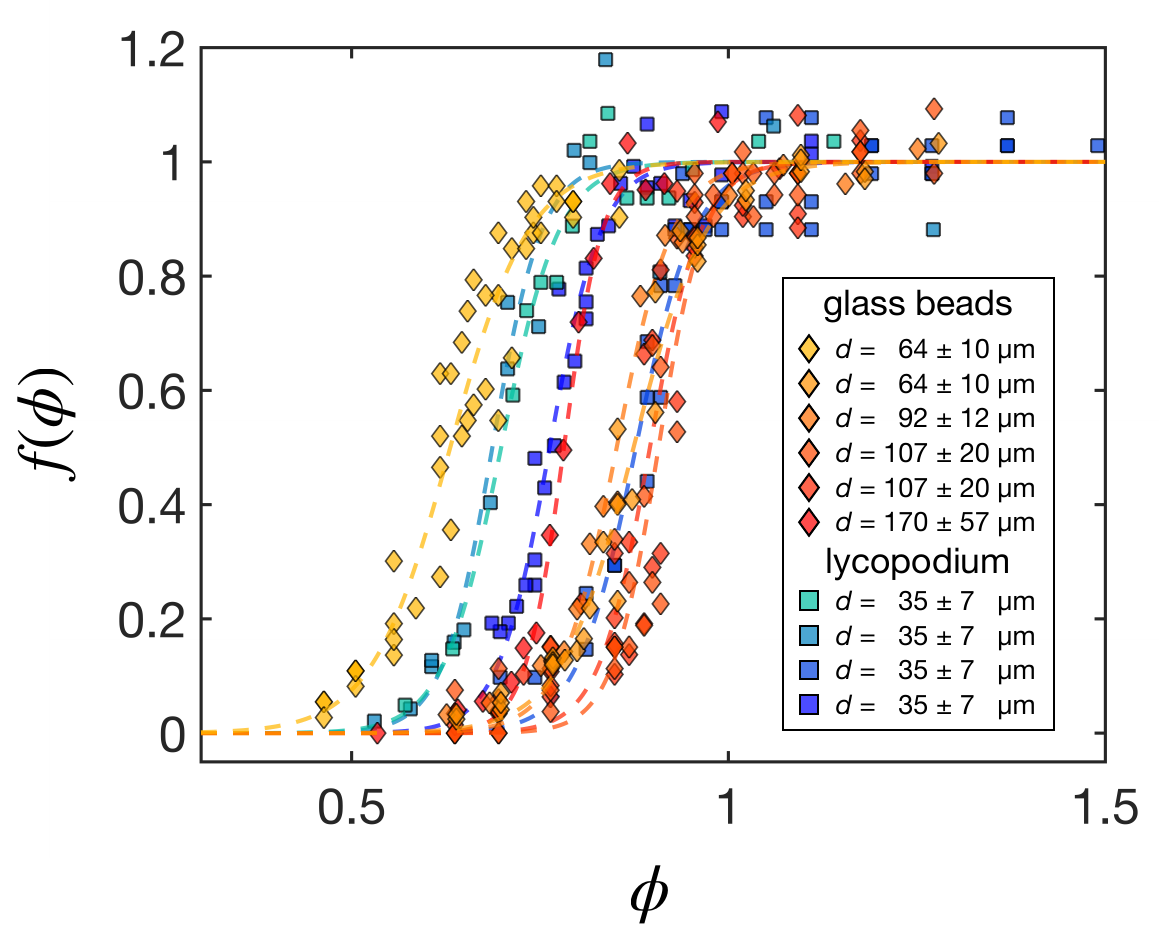}
	\caption{ $f(\phi)$ measured with glass beads and lycopodium rafts. The dashed lines represent the best fit with the logistic function Eq.~\ref{eq:f_th}. The values for $\beta$ and $\phi_c$ and $F_\mathrm{max}$ are listed in Table \ref{table_fit}. }
	\label{f_data}
\end{figure}

\textit{Appendix H: Raft measurements ---} The rafts are prepared by depositing a particle mass $m_g$ at the surface of a Langmuir--Blodgett trough of area $S_t$, giving a surface fraction $\phi = 3m_g/(2\rho_g d S_t)$. We use lycopodium particles with diameter $d = 35 \pm 7~\mu$m and effective density $\rho_g = 950$ kg/m$^3$~\cite{binks2005naturally}, and glass beads with diameters $d = 170 \pm 57$, $107 \pm 20$, $92 \pm 12$, and $64 \pm 10~\mu$m and density $\rho_g = 2600$ kg/m$^3$. The trough area is varied, and for each value of $S_t$ we measure the force $F_y$ exerted by the granular raft on the fiber. For each particle type, we find that $F_y(\phi)$ follows a logistic curve, $F_y(\phi)=F_\mathrm{max}/[1+\exp[-\beta(\phi-\phi_c)]]$, and fit the corresponding parameters $(F_\mathrm{max},\beta,\phi_c)$. We plot $f=F_y/F_\mathrm{max}$ as a function of $\phi$ in Fig.~\ref{f_data}, and as a function of the rescaled density $\beta(\phi-\phi_c)$ in Fig.~\ref{F4}. The fitted parameters are reported in Table~\ref{table_fit}.
\\

\begin{table}[h!]
\scalebox{0.85}{
\begin{tabular}{ p{3.5cm}||c|c|c|c}
\centering
 & $d$ ($\mu$m) & $F_\mathrm{max}$ ($\mu$N) & $\beta$  & $\phi_c$ (\%)\\
\hline
\textbf{lycopodium} marbles &32 $\pm$ 3   & -- & 15.4 $\pm$ 1.7    & 90 $\pm$ 1 \\
\hline
\textbf{lycopodium} raft   & 35 $\pm$ 7   & 109 $\pm$ 5 & 24.3 $\pm$ 5.2    & 76 $\pm$ 1 \\
                            & 35 $\pm$ 7   & 115 $\pm$ 4 & 33.3 $\pm$ 9.0    & 88 $\pm$ 1 \\
                            & 35 $\pm$ 7   & 87 $\pm$ 9 & 27.8 $\pm$ 12.8    & 69  $\pm$ 2 \\
                            & 35 $\pm$ 7   & 113 $\pm$ 7 & 32.8 $\pm$ 14.1    & 69  $\pm$ 2 \\
 \hline
\textbf{glass beads} raft  & 170 $\pm$ 57   &    198 $\pm$ 9   & 36.1 $\pm$    7.7    & 78 $\pm$ 1  \\
  &107 $\pm$ 20                             &    172 $\pm$ 21   & 27.3 $\pm$    10.9   & 89 $\pm$ 2 \\
  &107 $\pm$ 20                             &    205 $\pm$ 10   & 35.1 $\pm$    9.8    & 90 $\pm$ 1  \\
  &92 $\pm$ 12                              &    164 $\pm$ 9   & 27.9 $\pm$    6.7    & 84 $\pm$ 1  \\
  &64 $\pm$ 10                              &    153 $\pm$ 8   & 22.6 $\pm$    4.1    & 88 $\pm$ 1 \\
  &64 $\pm$ 10                              &    300 $\pm$ 23   & 19.8 $\pm$    4.8    & 63 $\pm$ 1 \\
\end{tabular}
}
	\caption{Logistic function fit parameters for lycopodium liquid marbles, rafts, and glass beads rafts.}
    \label{table_fit}
\end{table}

\textit{Appendix I: Fiber's effective spring constant ---} The fiber's effective spring constant $k$ is measured by the static deflection of the fiber under its own weight. The values obtained range from 10 mN/m to 60 mN/m and are in good agreement with the cantilever beam equation for small deformations, $k = 3 E I / L^3$, with $I = \pi d_f^4 / 64$ the fiber's area moment of inertia, and $E \approx$~70~GPa its Young’s modulus. We considered only the linear regime (Hooke's law), which provides a good approximation with an average error of less than 1\% and a maximum error of less than 7\% in the worst-case scenario.

\end{document}